\documentclass[twocolumn,showkeys,amsmath,amssymb]{revtex4}%
\usepackage{graphicx}
\usepackage{dcolumn}
\usepackage{bm}%
\usepackage{amsmath}%
\usepackage{amsfonts}%
\usepackage{amssymb}
\providecommand{\U}[1]{\protect\rule{.1in}{.1in}}
\begin{document}
\preprint{APS}
\title{Generalized versus\\selected descriptions of quantum $LC$-circuits}
\author{E. Papp}
\email{erhardt_papp_2005@yahoo.com}
\affiliation{Department of Theoretical Physics, West University of Timisoara, 300223, Romania}
\author{C. Micu}
\email{codrutamicu2004@yahoo.com}
\affiliation{Physics Department, North University of Baia Mare, 430122, Romania }
\author{O. Borchin}
\affiliation{Department of Theoretical Physics, West University of
Timisoara, 300223, Romania.}
\author{L. Aur}
\affiliation{Department of Theoretical Physics, West University of
Timisoara., 300223, Romania}
\date{\today}

\begin{abstract}
Proofs are given that the quantum-mechanical description of the $LC$-circuit
with a time dependent external source can be readily established by starting
from a more general discretization rule of the electric charge. For this
purpose one resorts to an arbitrary but integer-dependent real function $F(n)
$ instead of $n$. This results in a nontrivial generalization of the discrete
time dependent Schr\"{o}dinger-equation established before via $F(n)=n$, as
well as to modified charge conservation laws. However, selected descriptions
can also be done by looking for a unique derivation of the effective
inductance. This leads to site independent inductances, but site dependent
ones get implied by accounting for periodic solutions to $F(n)$ in terms of
Jacobian elliptic functions. Many-charge generalizations of quantum circuits,
including the modified continuity equation for total charge and current
densities, have also been discussed.

\end{abstract}
\maketitle


\section{Introduction}

A fundamental concept which is responsible for sensible effects in electronic
devices is the discreteness of the electric charge [1]. Quantizations of the
conductance in units of $e^{2}/hc$ [2,3], or of the magnetic flux in units of
$hc/e$ [4], can also be mentioned. Note that the charge of the electron is
$-e$, as usual. Handling the discretized charge also means that the
application of the discrete calculus, such as done by left ($\nabla$)- and
right-hand ($\Delta)$ discrete derivatives, is rather suitable. Looking for
explanations, we have to realize that looking for low dimensional nanoscale
systems on discrete spaces provide a deeper understanding of phenomena [5].
The same concerns discrete tight binding models relying naturally on
semiconductor quantum wells and nanoelectronic devices [1,6]. In latter cases
the coherence length gets larger than sample dimensions, which leads to
wealthy manifestations of quantum interference phenomena. Much progress has
also been done in the field of miniaturization of circuits. In this context it
has been found that the quantum mechanical description of $LC$ [7-10]-, $L$
[10,11]- and $RLC$ [12,13]-circuits can be done by resorting once more again
to the charge discretization. Studies in such fields are promising, as they
produce ideas for further technological developments.

We shall then use this opportunity to discuss in some more detail the
quantum-mechanical description of the mesoscopic $LC$-circuit with a time
dependent voltage source $V_{s}\left(  t\right)  $. So far the discrete
Schr\"{o}dinger-equation characterizing the $LC$-circuit has been established
by starting from the concrete charge eigenvalue equation [7,8,10,11]%

\begin{equation}
Q_{q}\left\vert n\right\rangle =nq_{e}\left\vert n\right\rangle \ \quad,
\end{equation}
where $Q_{q}$ denotes the Hermitian charge operator and where $n$ is an
integer playing the role of the discrete coordinate. Equation (1) shows, of
course, that the electric charge is quantized in units of the elementary
electric charge $q_{e}=e$. One could also say that $q_{e}=2e$ when dealing
with Cooper-pairs [14,15]. However, more general discrete
Schr\"{o}dinger-equations concerning $LC$-circuits can also be derived. For
this purpose we shall begin by performing a charge mapping like $Q_{q}%
\rightarrow\widetilde{Q}_{q}=F(Q_{q})$, where $F(n)$ is a real function of
$n$. Applying discrete derivatives to the eigenvalue equation of
$\widetilde{Q}_{q}$, i.e. to%

\begin{equation}
\widetilde{Q}_{q}\left\vert n\right\rangle =F(Q_{q})\left\vert n\right\rangle
=F(n)q_{e}\left\vert n\right\rangle \quad.
\end{equation}
results, surprisingly enough, in a generalized counterpart of the discrete
Schr\"{o}dinger-equation relying on (1), as well as in non-trivial
modifications of the charge conservation law.

The first task concerns the derivation of the canonically conjugated
observable, i.e. of suitable magnetic flux operators. We shall then obtain a
pair of non-Hermitian but conjugated magnetic flux operators. The product of
such operators is then responsible for the kinetic energy-term, i.e. for the
Hermitian operator of the square magnetic flux. Within the next stage of our
approach we shall look, however, for selected realizations of $F(n)$ yielding
a reliable description from the physical point of view. This proceeds by
establishing in a well defined manner the effective impedance for the quantum
circuit one deals with. Many-charge generalizations of quantum-circuit
equations can also be readily established.

\section{Preliminaries and notations}

We have to recall that the classical $RLC$-circuit is described by the balance equation%

\begin{equation}
L\frac{dI}{dt}+IR+\frac{Q}{C}=V_{s}\left(  t\right)  \ \quad,
\end{equation}
in accord with Kirchhoff's law, where the current, the inductance, the
capacitance and the resistance are denoted by $I=dQ/dt$, $L$, $C$ and $R$,
respectively. Periodic modulations of the voltage like $V_{s}(t)=V_{0}%
\cos(\omega t)$ are frequently used, in which case the circuit is
characterized by the impedance%

\begin{equation}
Z=R+i\left(  \omega L-\frac{1}{\omega C}\right)  \quad.
\end{equation}
Inserting, for convenience, $R=0$, leads to the Hamiltonian%

\begin{equation}
\mathcal{H}_{c}\left(  Q,\frac{\Phi}{c}\right)  =\frac{\Phi^{2}}{2Lc^{2}%
}+\frac{Q^{2}}{2C}-QV_{s}\left(  t\right)  \ \quad,
\end{equation}
where $\Phi=ILc$ stands for the magnetic flux. Accordingly, (3) is produced by
the Hamiltonian equations of motion%

\begin{equation}
I=\frac{dQ}{dt}=\frac{\partial\mathcal{H}}{\partial\left(  \Phi/c\right)
}=\frac{\Phi}{Lc}\ \quad,
\end{equation}
and%

\begin{equation}
\frac{d}{dt}\left(  \frac{\Phi}{c}\right)  =-\frac{\partial\mathcal{H}%
}{\partial Q}=-\frac{Q}{C}+V_{s}\left(  t\right)  \ \quad,
\end{equation}
as usual. This also means that the electric charge and the magnetic flux
divided by $c$, i.e. $Q$ and $\Phi/c$, are canonically conjugated variables.
This result suggest that the quantization of the $LC$-circuit should be done
in terms of the canonical commutation relation%

\begin{equation}
\left[  Q,\Phi\right]  =i\hbar c\ \quad,
\end{equation}
in which case one gets faced with the flux-operator [16]%

\begin{equation}
\Phi=-i\hbar c\frac{\partial}{\partial Q}\ .
\end{equation}
However, a such realization is questionable because the electric charge, such
as defined by (1) is not a continuous observable. This means that the
introduction of a discretized version of (9) like%

\begin{equation}
\Phi_{q}=-i\frac{\hbar c}{q_{e}}\Delta\quad,
\end{equation}
for which $\Phi_{q}^{+}=-i\hbar c\nabla/q_{e}$ is in order. The Hermitian
time-dependent Hamiltonian of the quantum $LC$-circuit can then be established as%

\begin{equation}
\mathcal{H}_{q}=\frac{\Phi_{q}^{+}\Phi_{q}}{2Lc^{2}}+\frac{\widetilde{Q}%
_{q}^{2}}{2C}-\widetilde{Q}_{q}V_{s}\left(  t\right)  \ \quad,
\end{equation}
in which $\mathcal{H}_{q}^{(0)}=\Phi_{q}^{+}\Phi_{q}/2Lc^{2}$ has the meaning
of the kinetic energy. This amounts to solve the discrete Schr\"{o}dinger-equation%

\begin{equation}
\mathcal{H}_{q}<n\mid\Psi(t)>=i\hbar\frac{\partial}{\partial t}<n\mid\Psi(t)>
\end{equation}
working, of course, within the charge-number representation. The Hermitian
momentum operator an also be readily introduced as $P_{q}=(\Phi_{q}+\Phi
_{q}^{+})/2$.

Note that right- and left-hand discrete derivatives referred to above proceed
as [17]%

\begin{equation}
\Delta f\left(  n\right)  =f\left(  n+1\right)  -f\left(  n\right)  \ =\left(
\exp(\partial/\partial n)-1\right)  f(n)
\end{equation}
and%

\begin{equation}
\nabla f\left(  n\right)  =f\left(  n\right)  -f\left(  n-1\right)  =\left(
1-\exp(-\partial/\partial n)\right)  f(n)\ \quad
\end{equation}
so that $\Delta^{+}=-\nabla$ and $\nabla\Delta=\Delta-\nabla\ $. In addition,
one has the product rule%

\begin{equation}
\nabla\left(  f\left(  n\right)  g\left(  n\right)  \right)  =g\left(
n\right)  \nabla f\left(  n\right)  +f\left(  n-1\right)  \nabla g\left(
n\right)  \ \quad,
\end{equation}
and similarly for $\Delta$.

\section{Generalized versions of the electric charge quantization}

Next let us apply discrete derivatives presented above both to (2) and
$\left\vert n\right\rangle $. One finds%

\begin{equation}
\widetilde{Q}_{q}\Delta=q_{e}F\left(  n+1\right)  \Delta+q_{e}\Delta F\left(
n\right)  \ \quad,
\end{equation}
and%

\begin{equation}
\nabla\widetilde{Q}_{q}=q_{e}F\left(  n-1\right)  \nabla+q_{e}\nabla F\left(
n\right)  \ .
\end{equation}
Performing the Hermitian conjugation gives $\nabla\widetilde{Q}_{q}%
=q_{e}F\left(  n\right)  \nabla$ and $\Delta\widetilde{Q}_{q}=q_{e}F\left(
n\right)  \Delta$, where $\widetilde{Q}_{q}^{+}=\widetilde{Q}_{q}$. Accordingly%

\begin{equation}
\left[  \widetilde{Q}_{q},\Delta\right]  =q_{e}\Delta F\left(  n\right)
\left(  1+\Delta\right)  \ ,
\end{equation}
and%

\begin{equation}
\left[  \widetilde{Q}_{q},\nabla\right]  =q_{e}\nabla F\left(  n\right)
\left(  1-\nabla\right)  \ .
\end{equation}

Now we are ready to introduce rescaled magnetic flux operators like%

\begin{equation}
\widetilde{\Phi}_{q}=-\frac{i\hbar c}{q_{e}}\left(  \frac{1}{\Delta F\left(
n\right)  }\Delta\right)  \ \quad,
\end{equation}
which can be viewed as the generalized counterparts of (10) and%

\begin{equation}
\widetilde{\Phi}_{q}^{+}=-\frac{i\hbar c}{q_{e}}\left(  \frac{1}{\nabla
F\left(  n\right)  }\nabla+\frac{1}{\Delta F\left(  n\right)  }-\frac
{1}{\nabla F\left(  n\right)  }\right)  \ .
\end{equation}
Accordingly, the interaction-free Hamiltonian is given by%

\begin{equation}
\mathcal{H}_{q}^{\left(  0\right)  }\rightarrow\widetilde{\mathcal{H}}%
_{q}^{\left(  0\right)  }=\frac{\widetilde{\Phi}_{q}^{+}\widetilde{\Phi}_{q}%
}{2Lc^{2}}\ \quad,
\end{equation}
which can be rewritten equivalently as%

\begin{equation}
\widetilde{\mathcal{H}}_{q}^{\left(  0\right)  }=-\frac{\hbar^{2}}%
{2\widetilde{L}\left(  n\right)  q_{e}^{2}}\left(  \widetilde{\Delta}%
-\nabla\right)  \ .
\end{equation}
This time the inductance gets rescaled as%

\begin{equation}
L\rightarrow\widetilde{L}\left(  n\right)  =L\left(  \nabla F\left(  n\right)
\right)  ^{2}\ \quad,
\end{equation}
whereas the discrete right hand derivative $\Delta$ is replaced
anisotropically by%

\begin{equation}
\widetilde{\Delta}=(1-G(n))\Delta\quad.
\end{equation}
One has%

\begin{equation}
G\left(  n\right)  =1-\left(  \frac{\nabla F\left(  n\right)  }{\Delta
F\left(  n\right)  }\right)  ^{2}\ \quad,
\end{equation}
which leads to sensible effects. Under such conditions the anisotropic
discrete Schr\"{o}dinger-equation for the single-charge amplitude
$C_{n}(t)=<n\mid\Psi(t)>$ is given by%

\begin{equation}
-\frac{\hbar^{2}(1-G(n))}{2\widetilde{L}\left(  n\right)  q_{e}^{2}}%
C_{n+1}\left(  t\right)  -\frac{\hbar^{2}}{2\widetilde{L}\left(  n\right)
q_{e}^{2}}C_{n-1}\left(  t\right)  +
\end{equation}%
\[
+\left[  \frac{\hbar^{2}}{\widetilde{L}\left(  n\right)  q_{e}^{2}}\left(
1-\frac{G\left(  n\right)  }{2}\right)  +\frac{q_{e}^{2}}{2C}F^{2}\left(
n\right)  -q_{e}F\left(  n\right)  V_{s}\left(  t\right)  \right]
C_{n}\left(  t\right)  =
\]%
\[
=i\hbar\frac{\partial}{\partial t}C_{n}\left(  t\right)  \ .
\]
which works in accord with (2) and (12). It is clear that (27) reproduces the
usual result [7]%

\begin{equation}
-\frac{\hbar^{2}}{2Lq_{e}^{2}}\nabla\Delta C_{n}(t)+\left(  \frac{q_{e}^{2}%
}{2C}n^{2}-q_{e}nV_{s}\left(  t\right)  \right)  C_{n}\left(  t\right)  =
\end{equation}%
\begin{align*}
=i\hbar\frac{\partial}{\partial t}C_{n}\left(  t\right)  \
\end{align*}
via $F(n)\rightarrow n$.

\section{Modified charge conservation laws}

One sees that (27), which differs in a sensible manner from (28), has a rather
complex structure such as involved by the $n$-dependence of coefficient
functions. Such structures exhibit a certain similarity to Schr\"{o}dinger
equations with a position dependent effective mass [18]. Furthermore, we have
to realize that (27) as it stands provides useful insights for more general
descriptions. Indeed, (27) produces a modified continuity equation like%

\begin{equation}
\frac{\partial}{\partial t}\rho_{n}(t)+\Delta J_{n}(t)=g_{n}(t)\quad,
\end{equation}
where%

\begin{equation}
\rho_{n}(t)=q_{e}\mid C_{n}(t)\mid^{2}\quad,
\end{equation}
denotes the usual charge density, whereas%

\begin{equation}
J_{n}(t)=\frac{\hbar}{\widetilde{L(}n)q_{e}}\operatorname{Im}\left(
C_{n}(t)C_{n-1}^{\ast}(t)\right)  \quad,
\end{equation}
stands for the related current density. The additional term in the continuity
equation is%

\begin{equation}
g_{n}(t)=q_{e}G(n)\frac{\widetilde{L}(n+1)}{\widetilde{L}(n)}J_{n+1}(t)\quad,
\end{equation}
which shows that there are additional effects which are able to affect the
time dependence of the charge density. This results in the onset of an extra
charge density like%

\begin{equation}
\rho_{n}^{(diff)}(t)=-G(n)\frac{\widetilde{L}(n+1)}{\widetilde{L}(n)}%
{\displaystyle\int\limits_{-\infty}^{t}}
J_{n+1}(t^{\prime})dt^{\prime}\quad,
\end{equation}
relying typically on the nonlinear attributes of the generalized charge
discretization function. The total charge density is then given by%

\begin{equation}
\rho_{n}^{(tot)}(t)=\rho_{n}(t)+\rho_{n}^{(diff)}(t) \quad,
\end{equation}
in which it has been assumed that $\rho_{n}^{(diff)}(t)\rightarrow0$ when
$t\rightarrow-\infty$.

\section{Introducing the effective impedance}

Equation (27) can also be interpreted in terms of an effective anisotropic
inductance by assuming three different realizations, namely $L_{1}%
(n)=\widetilde{L}(n)/\left(  1-G(n)\right)  $, $L_{2}(n)=\widetilde{L}(n)$ and
$L_{3}(n)=\widetilde{L}(n)/\left(  1-G(n)/2\right)  $. However, the isotropy
can be restored via%

\begin{equation}
L_{1}(n)=L_{3}(n)=\widetilde{L}(n)\quad,
\end{equation}
in which case%

\begin{equation}
G(n)=0\quad.
\end{equation}
Accordingly, one should have%

\begin{equation}
\nabla F(n)=\pm\Delta F(n)\quad,
\end{equation}
by virtue of (26), so that%

\begin{equation}
\nabla\Delta F(n)=F(n+1)-2F(n)+F(n-1)=0\quad,
\end{equation}
or%

\begin{equation}
F(n+1)=F(n-1)\quad,
\end{equation}
respectively. Under such conditions the modifications to the continuity
equation are ruled out, as one might expect.

Equation (38) has two kinds of solutions. First there is the linear realization%

\begin{equation}
F^{(1)}(n)=\alpha_{1}n+\beta_{1}\quad,
\end{equation}
where $\alpha_{1}$ and $\beta_{1}$ are parameters, for which the effective
inductance is $\widetilde{L}_{1}=\alpha_{1}^{2}L$. The rational charge
quantization is performed in terms of the fixings $\alpha_{1}=1/P$ and
$\beta_{1}=0$, where $P$ is a non-zero integer. Periodic functions with unit
period could eventually be considered. However, in such cases one has $\nabla
F(n)=\Delta F(n)=0$, which means in turn that such solutions can not be
accepted. Note that (40) produces sensible modifications going beyond (28).
Indeed, the wave function acquires an additional phase via%

\begin{equation}
C_{n}(t)\rightarrow C_{n}(t)\exp\left(  \frac{i}{\hbar}\beta_{1}q_{e}%
{\displaystyle\int\limits_{0}^{t}}
V_{s}(t^{\prime})dt^{\prime}\right) \quad,
\end{equation}
whereas the voltage is supplemented by an additional dc-component, as
indicated by the superposition%

\begin{equation}
V_{s}^{(1)}(t)=\alpha_{1}q_{e}n\left(  \frac{\beta_{1}q_{e}}{C}-V_{s}%
(t)\right)  \quad.
\end{equation}
This time the shifted harmonic oscillator term is given by%

\begin{equation}
V_{HO}(n)=\frac{q_{e}^{2}\alpha_{1}^{2}}{2C}n^{2}+\frac{q_{e}^{2}\beta_{1}%
^{2}}{2C}\quad,
\end{equation}
so that the total potential energy reads $V^{(tot)}=V_{s}^{(1)}(t)+V_{HO}(n)$.
Moreover, we are in a position to introduce the effective $n$-independent impedance%

\begin{equation}
\widetilde{Z}_{1}=i\left(  \omega\widetilde{L}_{1}-\frac{1}{\omega C}\right)
\quad,
\end{equation}
which proceeds in accord with (4), (24) and (40), where now $R=0$.

Equation (39) has to be solved in terms of periodic functions of double period
2, i.e. in terms o trigonometric and/or Jacobian elliptic functions. In the
former case we can propose the solution%

\begin{equation}
F^{(2)}(n)=\alpha_{2}\sin(\pi n+\delta_{2})+\beta_{2}\quad,
\end{equation}
producing an oscillatory charge, for which%

\begin{equation}
\nabla F^{(2)}(n)=2\alpha_{2}(-1)^{n}\sin\delta_{2}\quad.
\end{equation}
The corresponding $n$-independent effective inductance is given by
$\widetilde{L}_{2}=4\alpha_{2}^{2}\sin^{2}\delta_{2}$, so that the impedance,
say $\widetilde{Z}_{2}$, can be readily established in a close analogy with
(44). A further solution working in terms of Jacobian elliptic functions such
as given by%

\begin{equation}
F^{(3)}(n)=\alpha_{3}sn(2nK+\delta_{3})+\beta_{3}\quad,
\end{equation}
can also be proposed. Here $sn(u)$ denotes the sine amplitude, $u$ stands for
the argument, whereas $K=K(k)$ is the well known complete elliptic integral of
modulus $k$ [19]. Just note that $sn(-u)=-sn(u)$, $sn(u+4K)=sn(u) $ and
$sn(u+2K)=-sn(u)$. Now one has%

\begin{equation}
\nabla F^{(3)}(n)=-2\alpha_{3}sn(2nK+\delta_{3})\quad,
\end{equation}
which shows that this time one deals with the effective $n$-dependent inductance%

\begin{equation}
\widetilde{L}_{3}(n)=4\alpha_{3}^{2}sn^{2}(2nK+\delta_{3})L \quad,
\end{equation}
so that the same concerns the related impedance%

\begin{equation}
\widetilde{Z}_{3}(n)=i\left(  \omega\widetilde{L}_{3}(n)-\frac{1}{\omega
C}\right)  \quad.
\end{equation}
For convenience, we have restricted ourselves to periodic $F_{i}(n)$-functions
($i=2,3$) defined in terms of odd functions, as shown by (45) and (47). This
corresponds to the linear $\alpha_{1}n$-term in (40), but the flux dependence
of persistent currents in Aharonov-Bohm rings could also be invoked [20].
Further clarifications concerning this point remain, however, desirable.
Two-point impedances can also be established [21], which proceeds in terms of
eigenvalues of related Laplacian matrices.

\section{Many charge generalizations of quantum LC-circuits}

Many-charge generalizations of (11) like%

\begin{equation}
\mathcal{H}_{q}^{(MC)}=%
{\displaystyle\sum\limits_{j=1}^{N}}
\mathcal{H}_{q}^{(j)}%
\end{equation}
where%

\begin{equation}
\mathcal{H}_{q}^{(j)}=\frac{\Phi_{q}^{(j)+}\Phi_{q}^{(j)}}{2L_{j}c^{2}}%
+\frac{\widetilde{Q}_{q}^{(j)2}}{2\mathcal{C}_{j}}-\widetilde{Q}_{q}%
^{(j)}V_{s}\left(  t\right)  \
\end{equation}
can also be proposed. The charge operators, the inductances and the
capacitances are denoted by $\widetilde{Q}_{q}^{(j)}$, $L_{j}$ and
$\mathcal{C}_{j}$, respectively, where now $j=1,2,...,N$. Such Hamiltonians
are synonymous to many-body counterparts of (11). Accordingly, (2) gets
generalized as%

\begin{equation}
\widetilde{Q}_{q}^{(j)}\mid n_{j}>=q_{j}F_{j}(Q_{q}^{(j)})\mid n_{j}%
>=q_{j}F_{j}(n_{j})\mid n_{j}>
\end{equation}
where the $n_{j}$'s are integers which are responsible for the charge
eigenvalues. For the sake of generality, several charge scales, say $q_{j}$
instead of $q_{e}$, have also been assumed. The present charge eigenfunctions
are expressed by products like%

\begin{equation}
\mid n;N>=%
{\displaystyle\prod\limits_{j=1}^{N}}
\mid n_{j}>
\end{equation}
so that (53) gets reproduced as%

\begin{equation}
\widetilde{Q}_{q}^{(j)}\mid n;N>=q_{j}F_{j}(n_{j})\mid n;N>\quad.
\end{equation}
Accordingly, the flux operator relying on the $j$-charge is given by%

\begin{equation}
\Phi_{q}^{(j)}=-i\frac{\hbar c}{q_{j}}\Delta_{j}%
\end{equation}
in which case%

\begin{equation}
\Phi_{q}^{(j)+}=-i\frac{\hbar c}{q_{j}}\nabla_{j}%
\end{equation}
where this time $\Delta_{j}f(n_{j})=f(n_{j}+1)-f(n_{j})$ and $\nabla
_{j}f(n_{j})-f(n_{j}-1)$.

We then have to solve the discrete Schr\"{o}dinger-equation%

\begin{equation}
\mathcal{H}_{q}^{(MC)}C_{n;N}(t)=i\hbar\frac{\partial}{\partial t}C_{n;N}(t)
\end{equation}
by accounting for the factorization ansatz%

\begin{equation}
C_{n;N}(t)\equiv<n_{1},n_{2},...n_{N}\mid\Psi(t)>=%
{\displaystyle\prod\limits_{j=1}^{N}}
C_{n_{j}}^{(j)}(t)\quad.
\end{equation}
Having obtained single charge amplitudes via%

\begin{equation}
\mathcal{H}_{q}^{(j)}C_{n_{j}}^{(j)}(t)=i\hbar\frac{\partial}{\partial
t}C_{n_{j}}^{(j)}(t)
\end{equation}
opens the way to establish the $N$-charge amplitude in terms of (59). This
separation produces a unique solution if $N=2$. This means that (59) has to be
understood as a reasonable extrapolation of the well defined $N=2$-result
towards $N\eqslantgtr3$. The many charge version of (27) is then given by
\begin{equation}
-\frac{\hbar^{2}(1-G_{j}(n_{j}))}{2\widetilde{L}_{j}\left(  n_{j}\right)
q_{j}^{2}}C_{n_{j}+1}^{(j)}\left(  t\right)  -\frac{\hbar^{2}}{2\widetilde
{L}_{j}\left(  n_{j}\right)  q_{j}^{2}}C_{n_{j}-1}^{(j)}\left(  t\right)  +
\end{equation}%
\begin{align*}
+\left[  \frac{\hbar^{2}}{\widetilde{L}_{j}\left(  n_{j}\right)  q_{j}^{2}%
}\left(  1-\frac{G_{j}\left(  n_{j}\right)  }{2}\right)  +\frac{q_{j}^{2}%
}{2\mathcal{C}_{j}}F_{j}^{2}\left(  n_{j}\right)  -q_{j}F_{j}\left(
n_{j}\right)  V_{s}\left(  t\right)  \right]
\end{align*}
\begin{align*}
\cdot C_{n_{j}}^{(j)}\left(  t\right)  =i\hbar\frac{\partial}{\partial
t}C_{n_{j}}^{(j)}\left(  t\right)  \ .
\end{align*}
for $j=1,2,...,N$, where%

\begin{equation}
\widetilde{L}_{j}\left(  n_{j}\right)  =L_{j}\left(  \nabla_{j}F_{j}\left(
n_{j}\right)  \right)  ^{2}\ \quad
\end{equation}
and%

\begin{equation}
G_{j}\left(  n_{j}\right)  =1-\left(  \frac{\nabla_{j}F_{j}\left(
n_{j}\right)  }{\Delta_{j}F_{j}\left(  n_{j}\right)  }\right)  ^{2}\ \quad.
\end{equation}

Repeating the same steps as before leads to the generalized version of the
modified continuity equation%

\begin{equation}
\frac{\partial}{\partial t}\rho_{n_{j}}^{(j)}(t)+\Delta_{j}J_{n_{j}}%
^{(j)}(t)=g_{n_{j}}^{(j)}(t)
\end{equation}
\bigskip where%

\begin{equation}
\rho_{n_{j}}^{(j)}(t)=q_{j}\mid C_{n_{j}}^{(j)}(t)\mid^{2}%
\end{equation}

\begin{equation}
J_{n_{j}}^{(j)}(t)=\frac{\hbar}{\widetilde{L}_{j}(n_{j})q_{j}}%
\operatorname{Im}\left(  C_{n_{j}}^{(j)}(t)C_{n_{j}-1}^{(j)\ast}(t)\right)
\end{equation}
and%

\begin{equation}
g_{n_{j}}^{(j)}(t)=q_{j}G_{j}(n_{j})\frac{\widetilde{L}_{j}(n_{j}%
+1)}{\widetilde{L}_{j}(n_{j})}J_{n_{j}+1}^{(j)}(t)\quad.
\end{equation}

Our next step is to perform the $j$-summation in (64). This leads to the
derivation of total charge and current densities as%

\begin{equation}
\rho_{n}^{(tot)}(t)=%
{\displaystyle\sum\limits_{j=1}^{N}}
\rho_{n_{j}}^{(j)}(t)
\end{equation}
and%

\begin{equation}
\overrightarrow{J}_{n}^{(tot)}(t)=\left\{  J_{n_{1}}^{(1)}(t),J_{n_{2}}%
^{(2)}(t),...,J_{n_{N}}^{(N)}(t)\right\}
\end{equation}
respectively. Correspondingly, the continuity equation reads%

\begin{equation}
\frac{\partial}{\partial t}\rho_{n}^{(tot)}(t)+\overrightarrow{\Delta}%
\cdot\overrightarrow{J}_{n}^{(tot)}(t)=G_{n}^{(tot)}(t)
\end{equation}
where $\overrightarrow{\Delta}=\{\Delta_{1},\Delta_{2},...,\Delta_{N}\}$ and%

\begin{equation}
G_{n}^{(tot)}(t)=%
{\displaystyle\sum\limits_{j=1}^{N}}
g_{n_{j}}^{(j)}(t)\quad.
\end{equation}
The conservation of the total charge would then occur when $G_{n}%
^{(tot)}(t)=0$ irrespective of $t$. This happens if $F_{j}(n_{j}%
)\rightarrow\alpha_{j}n_{j}+\beta_{j}$, but the same concerns realizations
complying with (45) or (47), respectively. Alternatively, there are mutual
cancellation effects leading to $G_{n}^{(tot)}(t)=0$, which are worthy of
being considered in some more detail.

\section{Conclusions}

In this paper we succeeded to establish a more general quantum-mechanical
description of $LC$-circuits by starting from a generalized discretization
rule for the electric charge. To this aim one resorts to a real, but integer
dependent function $F\left(  n\right)  $ instead of $n$. This leads to the
generalized discrete Schr\"{o}dinger-equation (27), which reproduces the usual
result as soon as $F(n)=n$. A such generalized equation is able to incorporate
additional effects going beyond the charge conservation proceeding usually in
terms of ingoing and outgoing electron flows. Selected realizations of such
generalized descriptions are able to be done by resorting to additional
physical requirements. For this purpose we found it suitable to look for a
unique derivation of the effective inductance, as shown by (35). This leads to
a linear realization of $F(n)$ such as indicated by (40), but additional
periodic solutions with double period $2$ can also be proposed. Accordingly,
one gets faced both with trigonometric and elliptic solutions. Such solutions
are illustrated by (45) and (47), but other selections can also be done
specifically. One sees that the effective inductance remains independent of
$n$ both in terms of (40) and (45), but (49) exhibits clearly a non-trivial
$n$-dependence. Many charge generalizations of quantum $LC$-circuits can also
be done, as indicated in section 6. The modified continuity equation
concerning total charge and current densities has also been discussed.

It is worthy of being mentioned that the capacitance is sensitive to the
discreteness of the electronic charge, too [22]. This means that the
conduction is suppressed at low temperatures and small applied voltages, but
this phenomenon of \textquotedblleft Coulomb blockade\textquotedblright can be
removed by periodically modulated capacitive charging. Such effects have also
been discussed by resorting to pure capacity-design circuits [8].

The charge discretization is also able to serve to the quantum description of
other systems which are relevant in nanoelectronics, namely miniaturized $LC$
ladder-circuits. First steps along this direction have already been done, but
further investigations are in order [13, 23]. Such circuits contain cells
coupled capacitively, so that currents ($I_{n})$, voltages ($V_{n})$, charges
($Q_{n})$ and magnetic fluxes ($\Phi_{n})$ are site dependent. Within the
linear regime the current obeys the equation%

\begin{equation}
\nabla\Delta I_{n}=LC\frac{d^{2}I_{n}}{dt^{2}}\quad,
\end{equation}
and similarly for $V_{n}$. We have to remark that these equations are
equivalent to a linear Toda lattice [24]. However, nonlinearities may occur,
in which case $\Phi_{n}=Li_{0}f_{NL}(I_{n}/i_{0})$ instead of $\Phi_{n}%
=LI_{n}$, where $i_{0}$ denotes a current scale. This yields the modified equation%

\begin{equation}
\nabla\Delta I_{n}=i_{0}LC\frac{d^{2}f_{NL}(I_{n}/i_{0})}{dt^{2}}\quad,
\end{equation}
which shows that a site dependent inductance such as given by $L_{NL}(n)=L$
$f_{NL}(I_{n}/i_{0})$ has to be accounted for effectively. The same remains
valid for a nonlinear capacitance given by $C_{NL}(n)=Cg_{NL}(V_{n}/v_{0})$,
in which case%

\begin{equation}
\nabla\Delta V_{n}=v_{0}LC\frac{d^{2}g_{NL}(V_{n}/v_{0})}{dt^{2}} \quad,
\end{equation}
where $v_{0}$ is the voltage scale. We have to realize that the present
$n$-dependent inductance $\widetilde{L}(n)$ may be related or even identified
to $L_{NL}(n)$. So we found a possibility to handle charge and/or field
dependent parameters of the quantum $LC$-circuit in terms of corresponding
parameters of the nonlinear Toda-lattice. It is understood that time dependent
charges for which $I_{n}=dQ_{n}/dt\equiv d$ $\widetilde{Q}_{q}/dt$ should be
approached within the Heisenberg representation.

Preserving, however, both anisotropy and generality of (27), means that the
effective inductance is given by (24), so that the effective impedance is
$\widetilde{Z}(n)=i(\omega\widetilde{L}(n)-1/\omega C)$. Under such conditions
general nonlinear realizations of the charge discretization function $F(n)$,
although interesting from the mathematical point of view, are not easily
tractable. Indeed, they lead to position dependent hopping amplitudes, to
anharmonic effects as well as to complex valued energy dispersion laws.
Moreover, in such cases the equivalence between the $L$-ring circuit and the
electron on the 1D lattice under the influence of the induced time dependent
electric field is lost and the same concerns dynamic localization conditions
[10, 25]. However, there are reasons to emphasize that (27) is a promising
starting point towards applications concerning the complex motion of electrons
or of other carriers under modified charge conservation laws.

Unusual commutation relations like%

\begin{equation}
\left[  \widetilde{Q}_{q},\widetilde{\Phi}_{q}\right]  =-i\hbar c\left(
1+i\frac{q_{e}}{\hbar c}\Delta F(n)\widetilde{\Phi}_{q}\right) \quad,
\end{equation}
and%

\begin{equation}
\left[  \widetilde{Q}_{q},\widetilde{\Phi}_{q}^{+}\right]  =i\hbar c\left(
i\frac{q_{e}}{\hbar c}\nabla F(n)\widetilde{\Phi}_{q}^{+}-\frac{\nabla
F(n)}{\Delta F(n)}\right)  \quad,
\end{equation}
have also to be mentioned. Such relationships can be viewed as non-Hermitian
versions of generalized canonical commutation relations acting on
non-commutative spaces [26, 27], which looks rather challenging. Going back to
(40) yields, however, a closed algebra encompassing the kinetic energy, the
momentum and the charge, as indicated before [10].

Finally let us address the question of whether mesoscopic systems like quantum
circuits should lie definitely under the incidence of usual quantum
electrodynamics (QED) and of usual condensed matter theory or not. Strictly
speaking the answer seems to be negative. Indeed, being mesoscopic or
respectively nanosized is rather different from being a macroscopic condensed
matter system. First, the number of constituents is by now rather small.
However, the main point is that the miniaturization makes the coherence length
to be larger than the sample dimensions. This opens the way to the occurrence
of unexpected\ quantum interference phenomena, such as Aharonov-Bohm
oscillations of the conductance with respect to the external fields,
persistent currents, or Coulomb-blockade effects. It should be stressed that
such effects work in conjunction with the discreteness of the charge.
Moreover, there are parity dependent period doubling effects in the
oscillations of persistent currents in Aharonov-Bohm rings, but when such
rings are discretized only [5,28]. In addition, one deals with nontrivial
dynamic localization effects characterizing electrons under the influence of a
time dependent electric field [25] when the 1D line is replaced again by the
1D lattice. These latter effects show that the discreteness of the space has
to be accounted for, too. In other words, one deals specifically with new
physics relying on a new quantum phase [29], for which neither the
thermodynamic limit nor the ensemble averaging remain valid. Nevertheless,
signatures of the many-body Kondo effect are still able to be identified, such
as found before in the case of junctions between Aharonov-Bohm rings and leads
[30,31]. In other words we have to account for interplays between new and
former effects, as one might expect. Furthermore, the application of usual
relativistic QED to miniaturized composites looks rather unsuitable. This time
there are novel effects relying on non-local currents and quantum non-locality
[32] or on the advent of nonlinear relationships to the detriment of Ohm's law
[33], which prevent usual QED from being relevant to mesoscopic systems.

So we are in a position to realize that being discrete opens the way to a
deeper understanding of mesoscopic phenomena. The canonical quantization
should then be done by applying discrete derivatives instead of usual ones.
This results in a promising perspective towards a suitable description of
mesoscopic systems, now in terms developments provided by the application of
quantum mechanics to low dimensional systems on discrete spaces. The same
concerns, of course, the successful description of new phenomena
characterizing mesoscopic structures in terms of appropriate tight binding
hopping models.

\section*{Acknowledgments}

The authors are indebted to CNCSIS/Bucharest for financial support. \newpage

\section*{References}

\bibliographystyle{unsrt}

\end{document}